\documentclass[aps,twocolumn,groupedaddress,superscriptaddress,amsmath,amssymb,prb,longbibliography]{revtex4-2}
\usepackage{mathtools,amsmath,amsxtra}
\usepackage{amssymb}
\usepackage{physics}
\usepackage[english]{babel}
\usepackage{dsfont}
\usepackage{graphicx}
\usepackage{dcolumn}
\usepackage{bm}
\usepackage{bbold}
\usepackage[usenames,dvipsnames]{color}
\usepackage[colorlinks,citecolor=Blue,linkcolor=Red, urlcolor=Blue]{hyperref}
\usepackage{tikz}
\usetikzlibrary{decorations.pathmorphing}
\usetikzlibrary{arrows.meta}
\usepackage{braket}
\usepackage[normalem]{ulem}
\usepackage{lipsum}
\usepackage{caption}
\DeclareCaptionType{equ}[][]

\newcommand{\rosso}[1]{\textcolor{red}{#1}}

\begin{document}

\title{Proof-of-concept Quantum Simulator based on Molecular Spin Qudits} 

\author{S. Chicco}
\affiliation{Università di Parma, Dipartimento di Scienze Matematiche, Fisiche e Informatiche, I-43124, Parma, Italy}
\affiliation{UdR Parma, INSTM, I-43124 Parma, Italy}
\author{G. Allodi}
\thanks{These authors contributed equally.}
\affiliation{Università di Parma, Dipartimento di Scienze Matematiche, Fisiche e Informatiche, I-43124, Parma, Italy}
\author{A. Chiesa}
\thanks{These authors contributed equally.}
\affiliation{Università di Parma, Dipartimento di Scienze Matematiche, Fisiche e Informatiche, I-43124, Parma, Italy}
\affiliation{UdR Parma, INSTM, I-43124 Parma, Italy}
\affiliation{Gruppo Collegato di Parma, INFN-Sezione Milano-Bicocca, I-43124 Parma, Italy}
\author{E. Garlatti}
\thanks{These authors contributed equally.}
\affiliation{Università di Parma, Dipartimento di Scienze Matematiche, Fisiche e Informatiche, I-43124, Parma, Italy}
\affiliation{UdR Parma, INSTM, I-43124 Parma, Italy}
\affiliation{Gruppo Collegato di Parma, INFN-Sezione Milano-Bicocca, I-43124 Parma, Italy}
\author{C. D. Buch}
\affiliation{Department of Chemistry, University of Copenhagen, DK-2100 Copenhagen, Denmark}
\author{P. Santini}
\email{paolo.santini@unipr.it}
\affiliation{Università di Parma, Dipartimento di Scienze Matematiche, Fisiche e Informatiche, I-43124, Parma, Italy}
\affiliation{UdR Parma, INSTM, I-43124 Parma, Italy}
\affiliation{Gruppo Collegato di Parma, INFN-Sezione Milano-Bicocca, I-43124 Parma, Italy}
\author{R. De Renzi}
\email{roberto.derenzi@unipr.it}
\affiliation{Università di Parma, Dipartimento di Scienze Matematiche, Fisiche e Informatiche, I-43124, Parma, Italy}
\author{S. Piligkos}
\email{piligkos@chem.ku.dk}
\affiliation{Department of Chemistry, University of Copenhagen, DK-2100 Copenhagen, Denmark}
\author{S. Carretta}
\email{stefano.carretta@unipr.it}
\affiliation{Università di Parma, Dipartimento di Scienze Matematiche, Fisiche e Informatiche, I-43124, Parma, Italy}
\affiliation{UdR Parma, INSTM, I-43124 Parma, Italy}
\affiliation{Gruppo Collegato di Parma, INFN-Sezione Milano-Bicocca, I-43124 Parma, Italy}

\begin{abstract}
The use of $d$-level qudits instead of two-level qubits can largely increase the power of quantum logic for many applications, ranging from quantum simulations to quantum error correction. Molecular Nanomagnets are ideal spin systems to realize these large-dimensional qudits. Indeed, their Hamiltonian can be engineered to an unparalleled extent and can yield a spectrum with many low-energy states. In particular, in the last decade intense theoretical, experimental and synthesis efforts have been devoted to develop quantum simulators based on Molecular Nanomagnets. However, this remarkable potential is practically unexpressed, because no quantum simulation has ever been experimentally demonstrated with these systems. Here we show the first prototype quantum simulator based on an ensemble of molecular qudits and a radiofrequency broadband spectrometer. To demonstrate the operativity of the device, we have simulated quantum tunneling of the magnetization and the transverse-field Ising model, representative of two different classes of problems. These results represent an important step towards the actual use of molecular spin qudits in quantum technologies.
\end{abstract}                              
\maketitle

\twocolumngrid 

\indent 

Molecular Nanomagnets (MNMs), molecules whose magnetic core is typically made of one or few exchange coupled magnetic ions, have provided an ideal play ground to investigate fundamental phenomena, ranging from quantum tunneling of the magnetization in isolated molecules \cite{Mannini2011, Wernsdorfer1999} to hysteresis at 60-80 K of single-molecule origin \cite{Goodwin2017,Guo2018} or decoherence \cite{Hill,Takahashi2011}. A strength point of this class of materials is that their complex single-molecule spin dynamics can be accessed even by bulk measurements \cite{Baker2012,Chiesa2017}. Nevertheless, coherent manipulation and readout of a single TbPc$_2$ molecule was shown in a single-molecule transistor \cite{Thiele2014,Godfrin2017}.\\ Being controllable quantum objects, MNMs have attracted a considerable attention as qubits \cite{Gaitarev,Sessoli2019,Carretta2021}, thanks to the remarkable possibilities of engineering their Hamiltonian \cite{Timco2009} and the long coherence times (from hundreds of $\mu$s to ms) reported in Cu \cite{BaderNatComm14} or VO complexes \cite{Zadrozny2015,Atzori2016,SIMqubit}. Moreover, the possibility of controlling their quantum state by electric fields \cite{Fittipaldi2019,Liu2021} and the blueprint of a magnetic quantum processor \cite{Blueprint_} have been recently shown. These results are very interesting, but what makes MNMs really potentially disruptive for quantum technologies is the fact that they naturally provide multi-level quantum systems, i.e. qudits with large number of states \cite{Pineda2021,JPCLqec,Biard2021}. Indeed, the use of qudits as elementary units of computation \cite{Bullock2005,Anderson2015,Morvan2021,Asaad2020} 
can simplify or improve quantum algorithms \cite{Campbell2014,NatPhysToffoli,RevQudits,Hu2018,Weggemans2022,Fischer2023} and quantum sensing protocols \cite{Kristen2020}. Moreover, by enconding a protected qubit into a single multi-level object, quantum error correction could be implemented without the large overhead of resources required by qubit-based codes \cite{Pirandola2008,PRXGirvin,Hu2019,JPCLqec,npjQI,Mezzadri2023}.\\
In the last decade many efforts have been focused on using MNMs as quantum simulators (QSs) \cite{Santini2011,SciRepNi,modules,VO2,TacchinoQudits,Lockyer2022}. QSs are controllable quantum systems whose dynamics is externally driven in order to mimic the evolution of the "target" Hamiltonian, i.e. the Hamiltonian of the model that needs to be simulated. QSs made of molecular qudits would be very interesting, because problems involving quantum objects with many degrees of freedom can be solved more efficiently by going beyond the binary qubit logic. For instance, nuclear \cite{Rico2018} or bosonic \cite{TacchinoQudits} Hamiltonians can be naturally mapped to the higher dimensional qudit Hilbert space, avoiding the large growth of qubits \cite{DiPaolo2020} or complex gates \cite{Sawaya2020} typical of multi-qubit encodings. Moreover, a QS based on molecular qudits could embed quantum error correction. However, in spite of more than a decade of efforts, an experimental realization of a QS based on MNMs was still lacking, thus leaving their striking potential completely unexpressed \cite{Carretta2021}.\\
Here we show a working proof-of-concept quantum simulator based on an ensemble of $^{173}$Yb(trensal) MNM qudits  \cite{jacsYb} and we demonstrate its operation by implementing the quantum simulation of models representative of two different classes of problems: an integer spin $>1/2$ subject to quantum tunneling of the magnetization (QTM) and a pair of spins 1/2 coupled by Ising interaction in presence of a transverse field (TIM). In both cases our QS reproduces the correct physical behavior and the results are in good agreement with calculations. \\

\section{Quantum Hardware} 

The core of the quantum simulator consists of a crystal containing isotopically enriched [$^{173}$Yb(trensal)] molecules, doped at 1\% into its diamagnetic [Lu(trensal)] isostructural analogue (see Methods).
Due to the large crystal field splitting of Yb(III), each molecule behaves as an electronic spin qubit (effective spin 1/2) coupled to a 6-levels nuclear spin qudit $I=5/2$, providing $2 \times 6$ states. The corresponding spin Hamiltonian is given by:
\begin{eqnarray}\nonumber
H_0 &=& A_\parallel S_z I_z + A_\perp \left( S_x I_x + S_y I_y\right) + p I_z^2 \\
&+&  \mu_B \textbf{S} \cdot \textbf{g} \cdot \textbf{B}_0 + \mu_N g_I \textbf{I} \cdot \textbf{B}_0,
\label{eq:spinHam}
\end{eqnarray}
where the first two terms represent the strong axial hyperfine interaction ($A_\parallel = -898$ MHz, $A_\perp = -615$ MHz), the third one describes the nuclear quadrupolar coupling ($p=-66$ MHz) and the last two are the electronic ($g_x=g_y=2.9$, $g_z=4.3$) and nuclear ($g_I=-0.2592$) Zeeman terms. The parameters were determined in previous works \cite{Rollano2022,jacsYb} (see Supplementary Fig. 1). Static fields ${B}_0$ between 0.12 and 0.22 T are applied along $x$, orthogonal to the molecular $C_3$ symmetry axis (Fig. \ref{fig:Relax}-(a)). At these fields the electronic Zeeman energy is the leading term in \eqref{eq:spinHam}, thus the eigenstates are almost factorized and are labeled by the dominant electronic and nuclear spin components along $\textbf{B}_0$, $\ket{m_S,m_I}$. 
Here we focus on states $\ket{m_S=1/2,m_I}$, with $m_I=1/2,-1/2,-3/2,-5/2$ and use the simplified notation $\ket{0},\ket{1},\ket{2},\ket{3}$, as in Fig. \ref{fig:Relax}-(b). The corresponding transition frequencies are $f_1$ ($\ket{0}\leftrightarrow\ket{1}$, red), $f_2$ ($\ket{1}\leftrightarrow\ket{2}$, yellow) and $f_3$ ($\ket{2}\leftrightarrow\ket{3}$, blue). \\
The use of an ordered ensemble of identical qudits as QS has the advantage of yielding the expectation values with  high statistics directly in a single run. 
Full control of the qudits is achieved by addressing each energy gap using a flexible broadband NMR spectrometer equipped with a tailored multi-frequency probe spanning the frequency range.
The driving Hamiltonian is:
\begin{eqnarray}
    H_1(t) &=& \left( \mu_B g_z S_z + \mu_N g_I I_z\right) \\  \nonumber
    && \sum_m B_{1m} \sin(\omega_{m} t +\phi_{m}) \Theta(\tau_m/2 - |t-t_{0m}|) 
    \label{eq:driving}
\end{eqnarray}
where $\Theta$ is the Heaviside step function and the sum runs over different pulses of amplitude $B_{1m}$ (parallel to the $c$ axis), duration $\tau_m$, center $t_{0m}$, frequency $\omega_m/2\pi$ and phase $\phi_m$ addressing consecutive ($\Delta m_I = \pm 1$) transitions (i.e.,  $\omega_m = f_{\eta}\times 2 \pi$, with $\eta=1,2,3$).  
The simulator operates at 1.4 K, a temperature at which all the eigenstates are populated. Hence, we prepare an initial pseudo-pure state by proper sequences of pulses (see Sec. \ref{subsec:tunn} and Methods).\\

\section{Calibration} 

\begin{figure}[t!]
    \centering
    \includegraphics[width=0.48\textwidth]{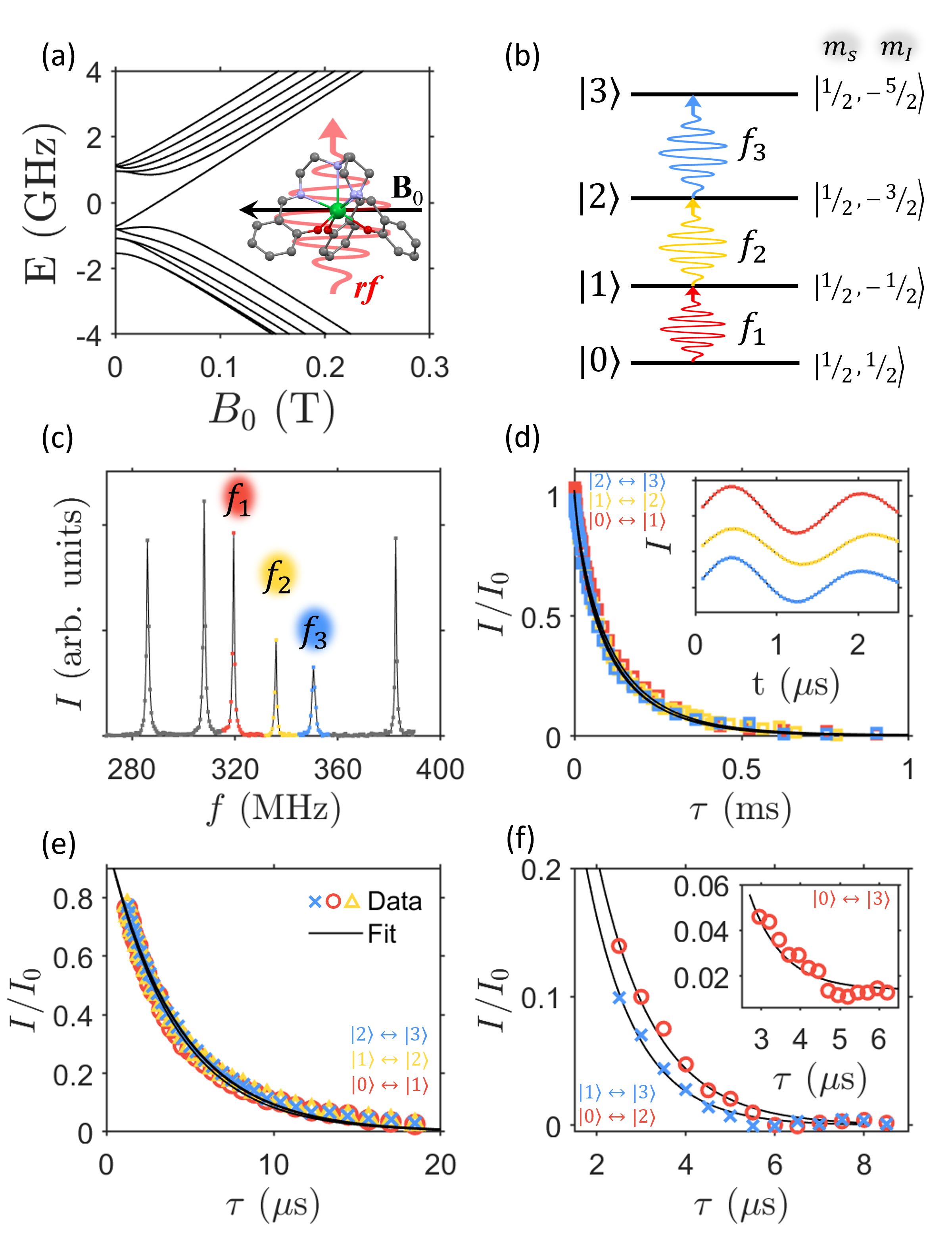}
    \caption{\textbf{Calibration of the Quantum Hardware.} (a) Calculated energy level diagram of [$^{173}$Yb(trensal)] with the static field $\textbf{B}_0$ perpendicular to the molecular $C_3$ axis. The molecule and the direction of the static (black) and driving (red) fields are shown as inset. (b) Scheme of the nuclear qudit subspace targeted in this work, with states labeled as $\vert 0 \rangle$, $\vert 1 \rangle$, $\vert 2 \rangle$ and $\vert 3 \rangle$ and transition frequencies as $f_1$, $f_2$ and $f_3$ in ascending order. (c) An example of the NMR spectrum of the [$^{173}$Yb(trensal)] qudit at $B_0=0.22$ T, with the peaks representing the nuclear transitions within the computational subspace  highlighted in colors. (d) Relaxation times $T_1^{\eta}$ measured (dots) on each of the nuclear transitions with the multi-frequency protocol, with $B_0=0.22$ T. Inset: coherent Rabi manipulation of the transitions indicated in panel (b) (labeled in color-code), demonstrating universal qudit control. (e) Phase memory time $T_2^{\eta}$ measured for each transition marked in panel (c), at $B_0=0.22$ T. (f) Double- (main) and triple- (inset) quantum coherence times. Error bars are within the size of the symbols.}
    \label{fig:Relax}
\end{figure}

We first need to show that a universal set of gates can be implemented in the QS and calibrate it.
The NMR spectrum is reported in Fig. \ref{fig:Relax}-(c), with the transition frequencies $f_1$ = 333.7 MHz, $f_2$ = 362.4 MHz and $f_3$ = 386.2 MHz highlighted in the corresponding color-code. [$^{173}$Yb(trensal)] has sharp spectral lines (FWHM $\sim$ 0.5 MHz), ensuring the possibility to individually address the transitions (see Fig. \ref{fig:Relax}-(b) and Supplementary Fig. 2).
To demonstrate full coherent control, we performed transient nutation experiments to induce $\Delta m_I = \pm 1$ Rabi oscillations with arbitrary phases between all the selected nuclear states (inset of Fig. \ref{fig:Relax}-(d)). These operations are the basic gates building up our quantum simulation sequences. These nutation experiments were also exploited to calibrate the duration of all the pulses at the working fields of the QS (see Supplementary Table S1).\\
Relaxation times much longer than the time needed to perform the full gate sequence and sufficiently long coherence times are required to perform a reliable quantum simulation. Thus, we measured all the relevant characteristic times $T_1^{\eta}$ and $T_2^{\eta}$ in the experimental conditions exploited in the quantum simulations. First, the relaxation times $T_1^{\eta}$ of the three selected transitions were probed by exploiting a double-frequency method. The signal decay is profiled by probing the transition $f_{\eta}$ between states $\ket{\eta-1}$ and $\ket{\eta}$ after an out-of-equilibrium surplus population is induced by an excitation pulse on the transition $f_{\eta \pm 1}$, to investigate the relaxation towards thermal equilibrium of diagonal elements of the density matrix (see Methods). The results obtained at the applied static field $B_0=0.22$ T are reported in Fig. \ref{fig:Relax}-(d), yielding $T_1^{\eta}$ values of the order of $200$ $\mu$s for all the transitions. Similar results were obtained at $B_0=0.12$ T (Supplementary Fig. 3).\\ 
Single-quantum coherence times $T_2^{\eta}$ (of superpositions between states with $\Delta m_I =1$) were measured by a standard Hanh-echo pulse sequence and are shown in Fig.\ref{fig:Relax}-(e) (see also Supplementary Fig. 4). The three transitions $f_\eta$ ($\eta=1,2,3$) show very similar $T_2^{\eta} \sim 8$ $\mu$s, significantly longer than simulation times. Additional key pieces of information for qudit-based architectures are the coherence times of superpositions involving $\Delta m_I >1$ states, the so-called multiple-quantum coherences. These superpositions are in fact created during quanum simulations and their characterization is therefore important for the design of optimized sequences. In order to extract multiple-quantum coherences, we first created the desired $\Delta m_I >1$ superposition exploiting $\pi$-pulses for state swaps (see Methods). After a variable delay, we used $\pi$ pulses to back swap the states and employ a $\frac{\pi}{2}-\frac{\pi}{2}$ sequence for detecting the decay of these coherences. Results for double- and triple-quantum coherences between the selected nuclear states are reported in Fig.\ref{fig:Relax}-(f) (main panel and inset, respectively). Since multiple-quantum superpositions involve states which are magnetically more different from each other, we found shorter coherence times with respect to single-coherences ($\sim 1.2$ $\mu$s for $\Delta m_I =2$ and $\sim0.7$ $\mu$s for $\Delta m_I =3$).\\ As shown by Figs. \ref{fig:QS_Ising} and \ref{fig:QS_Ising2}, these values permit the QS to capture the physics of the target models.\\

\section{Quantum Simulations}
The versatility of the QS is demonstrated by performing two different quantum simulations exploiting the multi-level structure of the molecular qudit: (i) the quantum tunneling of the magnetization of a single $S=1$ spin, where the $2S+1$ states of the target system are mapped onto the hardware levels and the unitary evolution is exactly decomposed into transitions between neighboring levels (Sec. \ref{subsec:tunn}).
(ii) The time dependence of the magnetization and of the correlation function for two spins 1/2 in a transverse magnetic field in two different regimes: either non-interacting or with an Ising coupling. Here the two-spin Hilbert space is mapped onto the single qudit energy levels and the unitary evolution induced by the target Hamiltonian is decomposed into a sequence of Suzuki-Trotter steps. This explores the possibility of encoding several spins into single qudits (see Sec. \ref{subsec:tim}).\\

\subsection{Quantum Tunneling}
\label{subsec:tunn}
\begin{figure}[t!]
    \centering
    \includegraphics[width=0.48\textwidth]{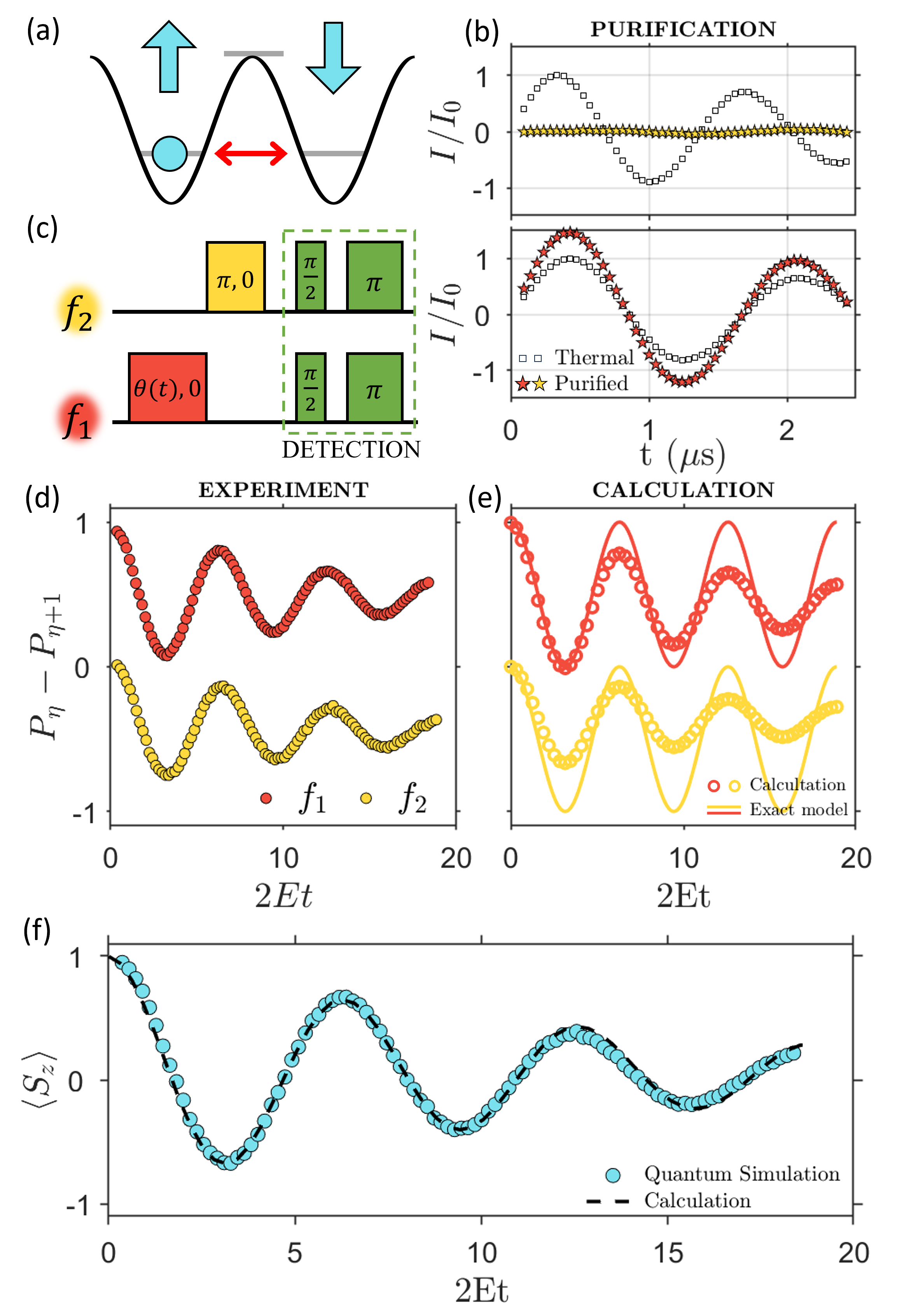}
    \caption{\textbf{Simulation of Quantum Tunneling of the Magnetization.} (a) Sketch of the double-well axial crystal field potential acting on a spin $S=1$ system prepared in $M=1$ (circle) and subject to quantum tunneling activated by rhombic anisotropic terms (red double-arrow). (b) Test of the purification protocol by sending pulses at frequency $f_2$ (top) and $f_1$ (bottom), respectively addressing $\ket{1}\leftrightarrow\ket{2}$ and $\ket{0}\leftrightarrow\ket{1}$ transitions, and comparing the driven dynamics before and after purification. (c) 2-frequency pulse sequence consisting 
    of a pulse of length $\theta(t) = Et$ at frequency $f_1$, followed by a $\pi$ pulse at frequency $f_2$ and concluded by Hahn-echo detection.
    (d) Difference of populations between consecutive levels $\ket{0}\leftrightarrow\ket{1}$ (red) and $\ket{1}\leftrightarrow\ket{2}$ 
    (yellow), measured at $B_0=0.12$ T by Hahn-echo sequences at frequencies $f_1$ and $f_2$, respectively, at the end of the quantum simulation. (e) Corresponding noiseless calculations (lines) or including measured single- and double-quantum $T_2^{\eta}$, as well as additional dephasing due to inhomogeneities of the driving field (circles). (f) Measured (blue circles) and calculated  (dashed line) expectation value of the magnetization of the target system. Error bars are within the size of the symbols.}
    \label{fig:QS_tunneling}
\end{figure}
We consider a $S=1$ target system characterized by the Hamiltonian (with $D>0$):
\begin{equation}
\mathcal{H}_{S} = -D S_z^2 + E \left( S_x^2-S_y^2 \right) .
    \label{eq:tunneling}
\end{equation}
For $E=0$, this corresponds to the double-well potential sketched in Fig. \ref{fig:QS_tunneling}-(a), where the ground state is a degenerate doublet with maximum absolute value of the magnetization (arrows in Fig. \ref{fig:QS_tunneling}-(a)), i.e. $M=\pm S$.
A small rhombic anisotropy term $E$ in $\mathcal{H}_S$ activates quantum tunneling through the barrier and hence a system prepared in one of the two wells oscillates between states with opposite magnetization. \\
To simulate the phenomenon, the three levels of the $S=1$ target system are mapped onto the hardware states $\ket{0}, \ket{1}$ and $\ket{2}$ of Fig. \ref{fig:Relax}-(b), which are initially in a thermal mixture because our experiment is not at $T=0$. Therefore, we prepare the initial pseudo-pure state in this subspace by first applying a $\pi/2$ pulse at frequency $f_2$ which creates a superposition between states $\ket{1}$ and $\ket{2}$ with equal amplitudes. This is followed by a waiting time $\sim 2.5 \; T_2^{2}$ to let the relative coherence decay. 
The resulting density matrix in the $\left\{ \ket{0},\ket{1},\ket{2} \right\}$ subspace is therefore of the form $\rho_{0-2} = \epsilon \ket{0} \bra{0} + (p_1+p_2)/2 \left( \ket{0} \bra{0} + \ket{1} \bra{1} + \ket{2} \bra{2} \right)$, with $\epsilon = p_0-(p_1+p_2)/2$ and $p_{\eta}$ the initial Boltzmann population of the energy states. Apart from normalization, this state is equivalent for quantum simulation to the {\it pure} density matrix $\rho_{0-2} = \ket{0} \bra{0}$. Indeed, the part of $\rho_{0-2}$ proportional to the identity in the considered subspace does not produce any signal in our experiment.
To check the "purification" procedure, we compare in Fig. \ref{fig:QS_tunneling}-(b) Rabi oscillations addressing transitions $\ket{1} \leftrightarrow \ket{2}$ (top) and $\ket{0} \leftrightarrow \ket{1}$ (bottom) before and after the sequence. Without purification (i.e., with thermal populations), a pulse of variable length at frequency $f_2$ induces oscillations between states $\ket{1}$ and $\ket{2}$. Conversely, after the purification sequence states $\ket{1}$ and $\ket{2}$ start with equal populations and hence Rabi oscillations are not observed (Fig. \ref{fig:QS_tunneling}-(b), top), as it would occur at $T=0$. Concerning the transition $\ket{0} \leftrightarrow \ket{1}$, the purification protocol enhances by about $50 \%$ their population difference, resulting in an amplification of Rabi oscillations (Fig. \ref{fig:QS_tunneling}-(b), bottom). In addition, we verified that coherences are lost after the waiting time (see Methods and Supplementary Fig. 5). \\
Having tested that the prepared state is spectroscopically equivalent to the pure state $\ket{0}$, we illustrate the simulation of the tunneling dynamics as a function of the simulation time $t$. The optimized sequence \footnote{The sequence is optimized for the quantum simulation of the tunneling of a $S=1$ spin starting from a $M=1$ state.} is shown in Fig. \ref{fig:QS_tunneling}-(c), ending with Hahn echo sequences at frequencies $f_1$ and $f_2$ to access differences between the populations of neighboring levels of the hardware $P_{\eta}-P_{\eta+1}$. \\
Results are shown in Fig. \ref{fig:QS_tunneling}-(d), in excellent agreement with calculations [Fig. \ref{fig:QS_tunneling}-(e)] including decoherence in Lindblad formalism (see Methods) and an additional decay ascribed to inhomogeneity of the driving field \cite{Chiorescu2007,jacsYb}. From these population differences we can extract the target observable $\langle S_z \rangle = P_0-P_2$,  i.e. the magnetization of the simulated system [see Fig. \ref{fig:QS_tunneling}-(f)]. This displays the expected quantum oscillation at frequency $E/\pi$, in very good agreement with calculations. \\

\subsection{Transverse Field Ising model}
\label{subsec:tim}
We now consider a different problem, represented by a target system of two spins 1/2, interacting via the Hamiltonian:
\begin{equation}
\mathcal{H}_{TIM} = b \left( s_{y1} + s_{y2} \right) + J s_{z1} s_{z2},
    \label{eq:TIM}
\end{equation}
where $s_{\alpha i}$ are spin 1/2 operators and we set $b=J$. 
\begin{figure}[t!]
    \centering
    \includegraphics[width=0.48\textwidth]{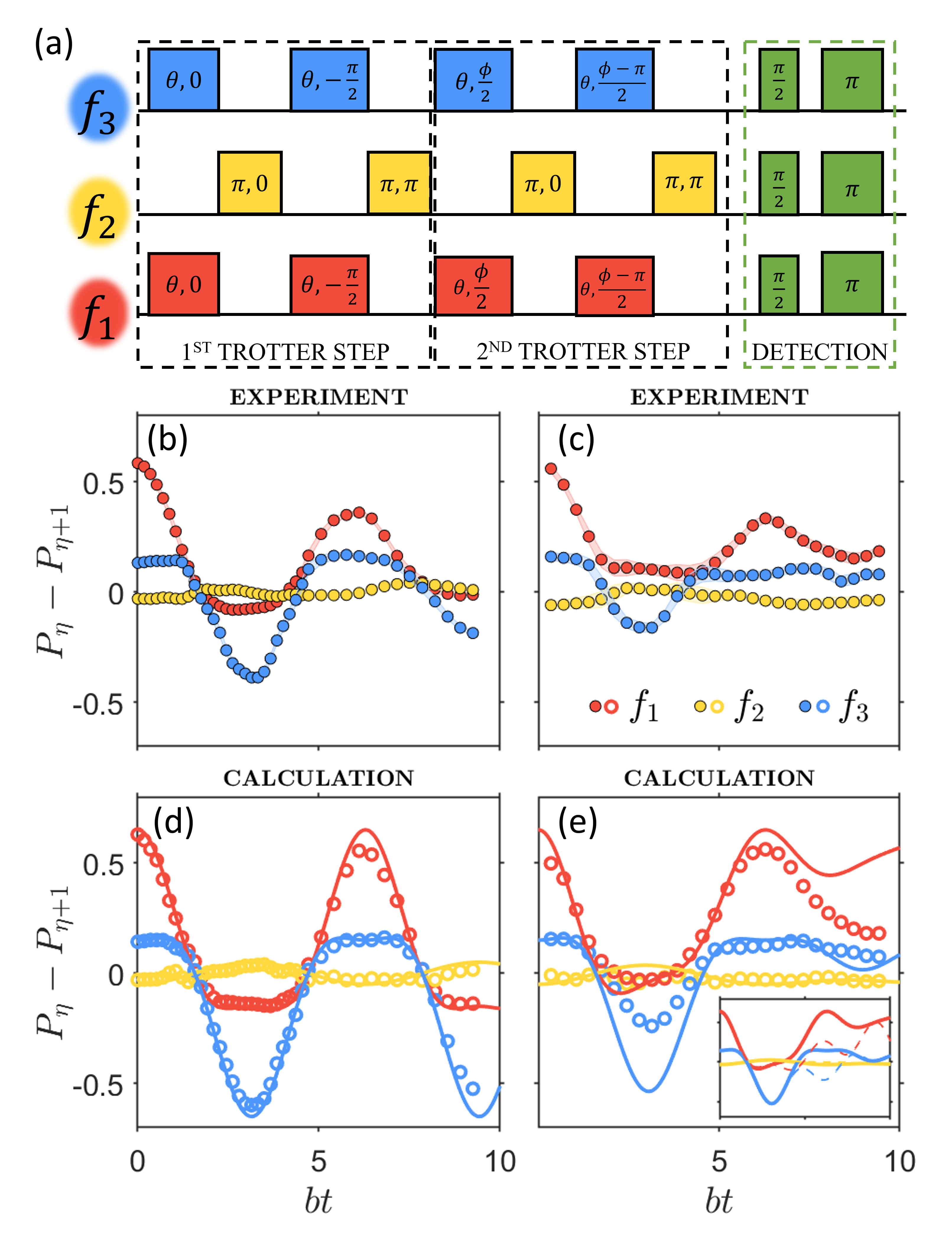}
    \caption{\textbf{Simulation of the Transverse Ising model.} (a) 3-frequency pulse sequence to implement the quantum simulation of the trasverse-field Ising model on 4 levels of the hardware qudit and to detect the final output. (b,c) Difference of populations between neighboring levels, measured at $B_0=0.22$ T by echo-sequences at the three driving frequencies $f_1$ (red), $f_2$ (yellow) and $f_3$ (blue) for the non interacting (b) and interacting (c) cases. The shaded areas represent the estimated experimental uncertainties in the amplitudes determination. (d,e) Corresponding calculations for $n=2$ with the inclusion of the incoherent Lindblad dynamics induced by the measured single-, double- and triple-quantum coherence times. Inset of panel (e): results for $n=2$ Suzuki-Trotter decomposition compared with the exact evolution induced by the target Hamiltonian (dashed lines).}
    \label{fig:QS_Ising}
\end{figure}
The quantum simulation of the corresponding time evolution $U(t) = e^{-i\mathcal{H}_{TIM}t}$ requires to decompose $U(t)$ into elementary operations which can be implemented on the hardware. In most qubit-based processors, this implies to separately simulate one- and two-body terms in Eq. \eqref{eq:TIM} and then to apply a Suzuki-Trotter (ST) approximation to $U(t)$, i.e.
\begin{equation}
U(t) \approx \left( e^{-i s_{z1} s_{z2} Jt/n} e^{-i \left( s_{y1} + s_{y2} \right) bt/n} \right)^n .
    \label{eq:Trotter}
\end{equation}
Such an approximation becomes exact for large number of Trotter steps $n$, at the price of an increasing number of noisy gates. Nevertheless, a proper trade-off can be found to reproduce the correct dynamics at not too large simulated times with a rather small $n$, thus limiting decoherence. \\
Here the four states of the target two-spin system $\left\{ \ket{\uparrow \uparrow}, \ket{\uparrow \downarrow}, \ket{\downarrow \uparrow}, \ket{\downarrow \downarrow}\right\}$ are mapped onto the qudit subspace $\left\{ \ket{0}, \ket{1}, \ket{2}, \ket{3}\right\}$. Hence, each one-body unitary gate in Eq. \eqref{eq:Trotter} is simulated by a pair of pulses of the same length $\theta=bt/n$ at frequencies $f_1$ and $f_3$, simultaneously addressing $\ket{0} \leftrightarrow \ket{1}$ and $\ket{2} \leftrightarrow \ket{3}$ transitions. This directly implements a rotation of the second qubit, i.e. $\exp [-i s_{y2} bt/n]$ \footnote{The rotation axis in the $xy$ plane is set by the phase $\phi_m$ of the pulse in Eq. \eqref{eq:driving}. $\phi_m=0$ implements a rotation about $y$.}. The same pulses, preceded and followed by a $\pi$ state-swap at frequency $f_2$, implement a rotation of the first qubit $\exp [-i s_{y1} bt/n]$. The resulting sequence yields the exact quantum simulation of $\mathcal{H}_{TIM}$ for the non-interacting ($J=0$ case) and it also corresponds to the first Trotter step of the interacting case (Fig. \ref{fig:QS_Ising}-(a), left). The simulation of the two-body term $\exp[-i s_{z1} s_{z2} Jt]$ on a qubit hardware would require controlled-phase gates at the end of each Trotter step. In our qudit architecture, this simply corresponds to adjusting phases
of the pulses addressing consecutive $\ket{0} \leftrightarrow \ket{1}$ and $\ket{2} \leftrightarrow \ket{3}$ transitions, as shown in  Fig. \ref{fig:QS_Ising}-(a) (for the second Trotter step). \\
An extension of the purification protocol illustrated above is used also in this second experiment to prepare the initial state (see Methods and Supplementary Figs. \rosso{6,7}). Detection of the output state is accomplished again by Hahn echo sequences at the frequencies $f_1$, $f_2$ and $f_3$. \\
\begin{figure}[t!]
    \centering
    \includegraphics[width=0.48\textwidth]{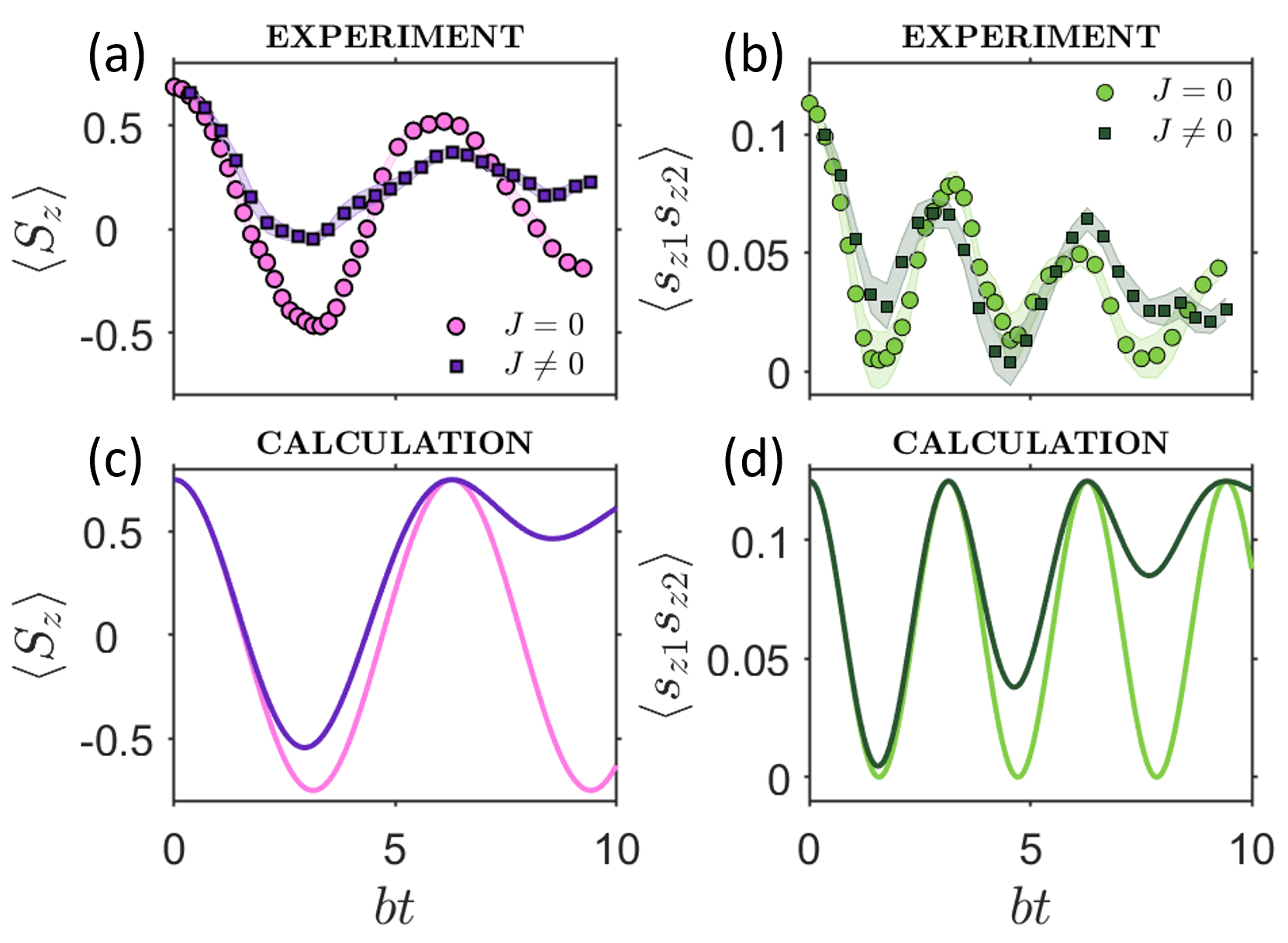}
    \caption{\textbf{Observables for the transverse-field Ising model.} Comparison between (a) the total magnetization $S_z = s_{z1}+s_{z2}$ and (b) the equal-time cross-correlation function $\langle s_{z1} s_{z2} \rangle$ for the examined two-spin model without ($J=0$) and with ($J=b$) Ising spin-spin coupling. Error bars represent the estimated uncertainties propagated from the experimental amplitudes of Fig. \ref{fig:QS_Ising}-(b,c). They are more important for $\langle s_{z1} s_{z2} \rangle$, where the signal results from a subtraction of experimental data. (c,d) Corresponding noiseless  calculations (lines) for $n=2$.}
    \label{fig:QS_Ising2}
\end{figure} 
Population differences measured at the end of the quantum simulation are reported in Fig. \ref{fig:QS_Ising} in non-interacting (b) and interacting (c) regimes, while corresponding observables are shown in Fig. \ref{fig:QS_Ising2}. Whereas for $J=0$ the simulation is exact, for $J \neq 0$ two Trotter steps are sufficient to capture the dynamics for $bt \lesssim 5$ (inset of Fig. \ref{fig:QS_Ising}-(e)). Nevertheless, we have explored also longer simulation times to make a more stringent demonstration of our capability of controlling the quantum hardware in presence of the complex dynamics induced by this sequence. 
Several of the pulses for the $J \neq 0$ case have been applied in parallel (Fig. \ref{fig:QS_Ising}-(a)) to make the duration of the sequences similar in the two cases and hence less dependent on decoherence.
The simulation could be extended to longer times by an exact decomposition in planar rotations, which however requires a significantly longer pulse sequence. \\
From Fig. \ref{fig:QS_Ising}-(b-d) we note a good agreement between experimental results (b,c) and calculations for $n=2$ (d,e), where the measured coherence times are included in a Lindblad formalism (circles). Pure dephasing here induces a damping of the oscillations of $P_{\eta}-P_{\eta+1}$ (dashed lines), but the non-trivial time dependence induced by the target Hamiltonian is well reproduced. Hence, our quantum simulator is able to catch the correct physical behavior of the target system. In particular, the total magnetization $S_z = s_{z1}+s_{z2}$ and the equal-time correlation $\langle s_{z1} s_{z2} \rangle$ simulated by the QS are reported in Fig. \ref{fig:QS_Ising2}-(a,b) and compared with exact calculations for $n=2$ (c,d). The QS predicts the oscillation frequency to be larger in the correlation than in the total magnetization, in good agreement with calculations. This agreement is remarkable especially for correlations, which are difficult to simulate because they are obtained from the difference of measured quantities (see Methods). In addition, the differences in the time-dependence between the interacting and non-interacting cases in the magnetization are captured by the QS.\\

\section{Scalability and perspectives}

We have demonstrated a proof-of-concept quantum device which explicitly makes use of the multi-level structure of Molecular Nanomagnets as a key resource for quantum simulation. This is done by following two different approaches, targeting different classes of problems: \\
\indent
1) the dynamics of a single multi-level system is directly mapped onto the energy levels of the qudit. This scheme can be  extended from $S>1/2$ problems to bosonic or fermionic degrees of freedom, which are of crucial interest but require complex encodings on multi-qubit platforms \cite{Rico2018,DiPaolo2020,Sawaya2020,TacchinoQudits}. \\
\indent
2) we have considered a multi-spin system whose Hilbert space is encoded into a single-qudit \cite{Carretta2021}. This approach is important for the scalability of the platform in the near future. By encoding several spins of the target Hamiltonian into the same qudit, we significantly reduce the number of two-body gates, which are usually the most error prone operations. Then, one can exploit a register consisting of several MNM (nuclear) qudits interacting via their electronic spins \cite{VO2}, to implement gates between different qudits. This can be still done in an ordered ensemble like a magnetically diluted crystal. \\
To further increase the scalability, the electronic spins can be used to activate an effective communication between distant qudits mediated by photons in superconducting resonators \cite{Blueprint_}, after having swapped quantum information from the nuclear spins. This is made possible by the specific choice of MNMs as elementary units. \\
The presence of metal ions whose spins are strongly coupled to nuclear ones provides specific features which 
make this architecture different from standard liquid-state NMR quantum computing (NMR-QC) \cite{Georgescu2014}. Indeed, besides being an important resource for scalability, this coupling can play a key role in specific protocols such as quantum-error correction \cite{JPCLqec,npjQI}. Moreover, it leads to large splittings between nuclear levels, 
making the thermal initialization in a pure state possible at mK temperatures. Finally, 
the unparalleled degree of tailoring of the spin Hamiltonian of MNMs \cite{modules} is a crucial advantage with respect to standard NMR-QC systems.\\ 
The next steps will involve the addition of higher-frequency pulses to control also electronic degrees of freedom, e.g., to mimic the interaction with a heat bath and then simulating open quantum systems \cite{Lockyer2022,Rogers2022}. Moreover, the use of more levels and/or multi-spin molecules will largely extend the class of Hamiltonians addressable by our Quantum Simulator.


\section{Methods}
\label{sec:methods}
\subsection{Synthesis}
A single crystal of isotopically enriched $^{173}$Yb(trensal) diluted at 1$\%$ into the isotructural Lu(trensal) was grown according to a published method for Er(trensal) \cite{Pedersen2014} where instead of using Er(OTf)$_3$·9H$_2$O as in the published method, $^{173}$Yb(OTf)$_3\cdot9$H$_2$O and Lu(OTf)$_3\cdot9$H$_2$O in the molar ratio 1:99 were used. Both Ln salts were synthesised according to a literature procedure, where the corresponding Ln$_2$O$_3$ was dissolved in boiling dilute triflic acid, and the Ln salt was obtained by slow evaporation of the corresponding solution (see Rev. Sci. Inst. 82, 096102 (2011)).  Isotopically enriched $^{173}$Yb$_2$O$_3$ was obtained from Neonest AB. 
Inductively coupled plasma mass spectrometry (ICP-MS) was used to determine the dilution of $^{173}$Yb(trensal) in Lu(trensal). ICP-MS was performed at the Department of Chemistry, University of Copenhagen on a Bruker Aurora Elite. Small crystals of $^{173}$Yb$_{0.01}$Lu$_{0.99}$(trensal) grown in the same tube as the one used for the experiments in the main text were dissolved in boiling nitric acid (14$\%$). The nitric acid was prepared by diluting TraceSelect grade conc. nitric acid with Milli-Q water. The solution was then diluted with TraceSelect grade nitric acid (2$\%$) until the concentration of $^{173}$Yb and Lu were within the calibration range of the instrument (1-50 ng/ml). Prior to determining the concentrations of $^{173}$Yb and Lu the ICP-MS instrument was tuned using six standard solutions with concentrations of Yb and Lu spanning the range 0-50 ng/ml. These standard solutions were prepared by diluting a reference solution from Inorganic Ventures using TraceSelect grade nitric acid (2$\%$). For the measurements of the Yb concentration the instrument was programmed only to detect the 173Yb isotope. The ICP-MS measurement afforded a ratio of 9:991 $^{173}$Yb:Lu.

\subsection{Apparatus}
The experimental apparatus for the characterization and control of the nuclear qudit has been specifically designed by combining the potentialities of the homemade broadband NMR spectrometer 'HyReSpect' \cite{Hyrespect} with a fast state-of-the-art Arbitrary Waveform Generator (Arb Rider AWG-5062\textbf{D}, hereafter AWG) from Active Technologies. The multi-frequency pulse sequences for the coherent manipulation of the nuclear qudit were in fact generated by the AWG externally triggered by spectrometer, while the spectrometer was devoted to the final state detection. The characteristics of the experimental setup are particularly suitable for the present experiment: a flat response over a wide frequency span, very short dead times ($<$1.3 $\mu$s) to make echo-detection compatible with the qudit phase memory time, fast RF switching, a broadband receiver stage and fast signal averaging. \\
The high sensitivity of the technique, enhanced by the strong hyperfine interactions of [$^{173}$Yb-trensal], allows the use of a NMR probe covering a wide frequency range ($\pm30$ MHz in our experiments), which can be be attained by inserting a parallel resistor in the LC circuit. The loss in sensitivity ($\propto \sqrt{Q}$) due to the diminished Q-factor of the probe was compensated by the isotopic enrichment of the target $^{173}$Yb species. 

\subsection{Calibration}
Rabi nutation experiments on each transition $f_{\eta}$ were performed by implementing a $(\theta(t))_{\eta}-(\pi)_{\eta}$ echo sequence, where the first pulse of variable length induces the nutation of the spin system in the rotating frame, while the refocusing is generated by the $\pi$-pulse. The decay observed in the intensity of Rabi oscillation (see inset of Fig. \ref{fig:Relax}-(d)) is dominated by the inhomogeneity of the driving field $B_1$, which adds to the $1/T_2^{\eta}$ rate (see Sec. \ref{subsec:calc} below).\\
Relaxation times $T_1^{\eta}$ between each pair of levels were measured by exploiting a double-frequency sequence generated by the AWG, of the type $(\pi)_{\eta \pm 1}-\tau-(\frac{\pi}{2})_{\eta}-(\pi)_{\eta}$. Indeed, the sequence to measure the time $T_1^{\eta}$ (corresponding to the transition $f_{\eta}:\ket{\eta-1}\leftrightarrow\ket{\eta}$) consists of (i) a population transfer to one of the two targeted nuclear states induced by $\pi$-pulse on a neighboring transition $f_{\eta \pm 1}$, (ii) the detection of the increment of the Hanh-echo signal on $f_{\eta}$ due to the induced out-of-equilibrium surplus population. The variable delay $\tau$ enables the determination of time required for the recovery of the thermal state populations on the targeted nuclear states $\ket{\eta-1}$ and $\ket{\eta}$, i.e. $T_1^{\eta}$. The $T_1^{\eta}$ decays are then subtracted by the Hahn-echo initial amplitude of the transition used for the detection.\\
Single-quantum coherence times $T_2^{\eta}$ were measured by a standard $(\frac{\pi}{2})_{\eta}-\tau-(\pi)_{\eta}$ Hanh-echo sequence, exploiting the standard spectrometer setup. The measurement of the multiple-quantum coherences required instead a multi-frequency pulse sequence generated by the AWG, for the preparation of the desired double- or triple-coherent superposition of states by addressing only consecutive transitions. The sequence for the double-quantum coherences can be written as: $(\pi/2)_{\eta+1}-(\pi)_{\eta+2}-\tau-(-\pi)_{\eta+2}$. First, a coherent superposition $\alpha\ket{\eta}+\beta\ket{\eta+1}$ is created between consecutive states by addressing the transition $f_{\eta+1}$. A $\pi$-pulse on $f_{\eta+2}$ is then used to implement a state-swap between $\ket{\eta+1}$ and $\ket{\eta+2}$, yielding the desired double-quantum coherent superposition $\alpha\ket{\eta}+\beta\ket{\eta+2}$. After a variable delay $\tau$ to follow the coherence decay, a $(-\pi)$ pulse on $f_{\eta+2}$ is implemented to back-swap the states. This final step recovers the now-decayed single-quantum coherent superposition on $f_{\eta+1}$, which can be detected by the spectrometer. For triple-quantum coherences, an additional $(\pi)_3$ pulse (together with the corresponding back-swap $(-\pi)_3$ one) is needed in order to prepare the $\alpha\ket{0}+\beta\ket{3}$ coherent superposition.\\
Multiple-quantum coherences were then measured by exploiting a $(\frac{\pi}{2})_{\eta+1}-(\frac{\pi}{2})_{\eta+1}$ detection sequence, where the first pulse was generated by the AWG and the last one by the spectrometer (hence only the latter was phase-coherent with the detection reference). The spin coherence induced by the first $\frac{\pi}{2}$ pulse, which would appear in principle as a (not observable) spin echo, is also encoded by this pulse into population differences. Such a longitudinally encoded frozen-in replica of the phase coherence present after the first pulse is then  turned into transverse coherence by the second $\frac{\pi}{2}$ pulse and then detected by the spectrometer as a "stimulated spin echo", as this process is referred to in the NMR literature. We stress that the detected signal cannot be due to either the trivial Hahn echo of the two $(\frac{\pi}{2})_{\eta+1}$ pulses themselves, nor any other combination of pulses generated  by the AWG alone. Since the spectrometer and the AWG are mutually incoherent, such spin echoes would average out on signal accumulation. On the contrary, reciprocal coherence of the two instruments is not need if spin coherence is first encoded in populations, as sketched above. The same detection method was used to measure the decay of the coherences induced by the pseudo-purification sequences, to check that they are completely lost after the waiting time $\sim 2.5 \; T_2^{\eta}$ before starting the quantum simulation (see Supplementary Figs. 5,7). For the quantum simulation of the Transverse Field Ising model, the pseudo-pure state was prepared with a $(\pi)_3-(\frac{\pi}{2})_2$ sequence. The first $\pi$ pulse induces a state-swap between $\ket{2}$ and $\ket{3}$, followed by the $\frac{\pi}{2}$ on $f_2$ creating a superposition between states $\ket{1}$ and $\ket{2}$ with equal amplitudes. Given the very similar Boltzmann population differences of the three involved levels, this sequence yields (apart from a contribution proportional to identity and a scale factor) a dominant population in $\ket{0}$ (0.75), small populations in $\ket{1}$ (0.11) and $\ket{2}$ (0.14). This enabled us to test the simulation starting from a non trivial initial state.\\
Quantum simulations were performed with an oscillating field $B_1\sim1$ G and $B_1\sim5$ G (depending on the addressed frequency) for the Quantum Tunneling Hamiltonian and for the Transverse Ising model, respectively.\\
All the detected echoes were then Fourier-transformed, phase-corrected and analyzed in the frequency domain by picking the spectral amplitude of the echo at a fixed frequency shift. 

\subsection{Observables}
The Hahn echo sequences at the end of the quantum simulations measure the differences between the populations of neighboring levels of the hardware $P_{\eta}-P_{\eta+1}$. From these quantities it is possible to extract physical observables.\\ 
For the quantum tunneling problem, we extracted the observable:
\begin{equation}
\langle S_z \rangle = [(P_0 - P_1) + (P_1 - P_2)] = P_0-P_2,
\end{equation}
i.e. the magnetization of the simulated system, reported in Fig. \ref{fig:QS_tunneling}-(f). The same quantity  $\langle S_z \rangle$ was extracted for the Transverse Ising model (see Fig. \ref{fig:QS_Ising2}-(a)), as 
\begin{equation}
\langle S_z \rangle = [(P_0 - P_1) + (P_1 - P_2) + (P_2 - P_3)] = P_0-P_3.
\end{equation}
For this Hamiltonian we have also extracted the equal-time correlation $\langle s_{z1} s_{z2} \rangle = \frac{1}{4}[(P_0-P_1) - (P_3-P_4)]$, shown in Fig. \ref{fig:QS_Ising2}-(b)).  

\subsection{Numerical calculations}
\label{subsec:calc}
Numerical calculations to reproduce the implemented quantum simulations have been performed by solving the Lindbald master equation:
\begin{equation}
\dot{\rho} = -\frac{i}{\hbar} [H,\rho] + \sum_{\eta \eta'} \gamma_{\eta \eta'} \rho_{\eta \eta'} \ket{\eta} \bra{\eta'} , 
    \label{eq:lindblad}
\end{equation}
where $\rho$ is the system density matrix in the eigenbasis, $\rho = \sum_{\eta \eta'} \rho_{\eta \eta'} \ket{\eta} \bra{\eta'}$, $H=H_0+H_1(t)$ is the system Hamiltonian (including time-dependent pulses) and $\gamma_{\eta \eta'}$ are pure dephasing rates of each specific superposition between eigenstates $\ket{\eta}$ and $\ket{\eta'}$. 
In the reported experiments $\ket{\eta} \approx \ket{m_S, m_I}$ and we have focused on the subspace with fixed $m_S=1/2$. 
Hence, rates $\gamma_{\eta \eta'}$ between states with different $m_I$ correspond to the inverse of the single and multiple-quantum coherence times discussed in the main text. Additional mechanisms depending on the details of the setup, like  inhomogeneities of the driving fields, could contribute to $\gamma_{\eta \eta'}$. These additional dephasing rates have been determined in the quantum tunneling experiment from the observed damping of the oscillations and included in the corresponding calculations. Conversely, to pinpoint the effect of decoherence in the complex dynamics associated with the TIM model, only the measured $T_2^{\eta}$ (single- and multi-quantum) have been included in the calculations.
The detection procedure has also been simulated. We have found that here pure dephasing acts practically as an overall scaling factor on the measured signal. Hence, we have re-scaled both signal and calculations to the known value at $t=0$. \\

\subsection{Sequence optimization}
The quantum simulation of the TIM model (target Hamiltonian \eqref{eq:TIM}) involves a Suzuki-Trotter decomposition in which rotations of the target qubits are alternated to an entangling ZZ evolution $U_{ZZ}(J\tau) = \exp[-i s_{z1} s_{z2} J \tau]$, $\tau=t/n$.
In order to reduce the number of pulses to be subsequently implemented, we have exploited the following identity:
\begin{equation}
R_{y}^{(1)} (\beta) R_{y}^{(2)} (\beta) U_{ZZ}(\alpha) = U_{ZZ}(\alpha) R_{c}^{(1)}(\beta,\alpha) R_{c}^{(2)}(\beta,\alpha), 
    \label{eq:sequence}
\end{equation}
where $R_{y}^{(i)}(\beta) = \exp[-i s_{yi} \beta]$, $R_{c}^{(1)}(\beta) = R_\alpha^{(1)}(\beta) \otimes \ket{0}\bra{0} + R_{-\alpha}^{(1)}(\beta) \otimes \ket{1}\bra{1}$ and $R_\alpha(\beta) = \exp[ -i ( \cos \alpha \; s_{y} - \sin \alpha \; s_x )\beta ]$. Analogous expressions hold for $R_{c}^{(2)}(\beta)$. In practice this corresponds to including $U_{ZZ}$ in the subsequent planar rotation. The rotation axis in the plane ($\alpha$) corresponds to the phase factor of the pulse.
Note that the effect of the entangling $U_{ZZ}$ gate is still present, because $R_{c}^{(i)}$ are conditional (entangling) gates in the two-qubit basis of the target system.\\

\section{Acknowledgements}
This work received financial support from European Union – NextGenerationEU, PNRR MUR project PE0000023-NQSTI, from the European Union’s Horizon 2020 program under Grant Agreement No. 862893 (FET-OPEN project FATMOLS), from the Novo Nordisk foundation under grant NNF21OC0070832 in the call "Exploratory Interdisciplinary Synergy Programme 2021" and from Fondazione Cariparma.

\section{Authors contributions}
S.Ch. and G.A. set up and performed the experiments designed by A.C., E.G. and S.C., after discussions with P.S. and R.DR. Analysis of the experimental results was carried out by S.Ch., G.A. and R.DR. Numerical
calculations were performed by A.C. and E.G.. Isotopically enriched crystals were prepared by C.D.B. and S.P.
P.S., R.DR., S.P. and S.C. conceived the work. A.C., E.G. and S.C. wrote the manuscript with
input from all co-authors.

\end{document}